\documentclass{article}

\usepackage{graphicx} 
\usepackage{booktabs} 
\usepackage[font=small,labelfont=bf]{caption} 
\usepackage{amsfonts, amsmath, amsthm, amssymb} 
\usepackage{enumerate}

\newtheorem{Thm}{Theorem}
\newtheorem{Def}{Definition}

\title{Are numerical theories irreplaceable? \\A computational complexity analysis.}
\date{}

\begin{document}
\maketitle
\begin{center}
\noindent \textbf{Nickolay Vasiliev$^{1,2}$} {(\texttt{vasiliev@pdmi.ras.ru})}\ \\ 
\textbf{Dmitry Pavlov$^3$} { (\texttt{dpavlov@iaaras.ru})}\\[0.5cm] 
\end{center}
\noindent  $^1$ St. Petersburg Department of V.\,A.\,Steklov Institute of Mathematics \\
       $^2$ St. Petersburg Electrotechnical University  \\
$^3$ Institute of Applied Astronomy RAS

\bigskip

\noindent {\footnotesize{Presented in a form of a poster at the Seventh International Meeting on Celestial Mechanics
(CELMEC VII) on 5 September 2017 at San Martino al Cimino, Italy.}}

\begin{abstract}

It is widely known that numerically integrated orbits are more precise
than analytical theories for celestial bodies. However, calculation
of the positions of celestial bodies via numerical integration
at time $t$ requires the amount of computer time proportional to $t$,
while calculation by analytical series is usually asymptotically faster.
 
The following question then arises: can the precision of numerical theories
be combined with the computational speed of analytical ones? We give
a negative answer to that question for a particular three-body problem
known as Sitnikov problem.

A formal problem statement is given for the the initial value problem (IVP) for
a system of ordinary dynamical equations. The computational complexity of this
problem is analyzed. The analysis is based on the result of Alexeyev (1968--1969)
about the oscillatory solutions of the Sitnikov problem that have chaotic
behavior. We prove that any algorithm calculating the state of the dynamical
system in the Sitnikov problem needs to read the initial conditions with
precision proportional to the required point in time (i.e. exponential in
the length of the point's representation). That contradicts the existence
of an algorithm that solves the IVP in polynomial time of the length
of the input.
\end{abstract}

\def\xvec{\mathbf{x}}
\def\fvec{\mathbf{f}}
\def\R{\mathbb{R}}
\def\Q{\mathbb{Q}}
\def\N{\mathbb{N}}
\def\Z{\mathbb{Z}}
\def\d{\mathrm{d}}
\def\qvec{\mathbf{q}}
\def\pvec{\mathbf{p}}
\def\vvec{\mathbf{v}}
\def\P{\mathcal{P}}
\def\LENGTH{\textrm{LENGTH}}

\section*{Introduction}

Analytical theories of planets (Brumberg 1991, Simon et al. 2013)
and the Moon (Chapront and Francou 2003, Ivanova 2014) allow
to determine the positions of bodies using series over the time parameter
$t$. Technically, those series work on arbitrarily large $t$, requiring
no more than $\P(\LENGTH(t))$ of computation time, where $\P$ is some polynomial
and $\LENGTH(t)$ is the length of $t$'s machine representation. For
fixed-precision $t$, $\LENGTH(t)$ is $\lceil\log t\rceil$ plus constant.

Conversely, planetary and lunar theories used in modern applied astronomy
(Pitjeva 2015, Folkner et al. 2014, Fienga et al. 2014) are based on numerical
integration, which requires at least $O(t)$ time to calculate the positions of
bodies at time $t$, regardless of precision.

Numerical theories are used because their accuracy matches the precision
of modern astronomical observations (radio observations of spacecraft,
lunar laser ranging). Analytical theories fail to provide that level
of accuracy. More to say, they are often fit to the existing numerical
theories rather than to observations.

Facing the problem of analyzing the behavior of the Solar system
for billions of years into the future or into the past, one could
wonder whether it is possible to use analytical theories in order
to get the $O(\P(\log t))$ computation time factor instead of $O(t)$.
At this point, one can not ignore the phenomenon of dynamical chaos:
high sensitivity of a dynamical system to initial conditions in presence
of quasi-periodic orbits.

Clearly, existing analytical theories are not purposed to calculate
chaotic orbits. In this work, we take a more general view on the
question of the computational complexity of the initial value
problem, not bound to any particular algorithm of computation.

In the next sections, a very short review is given about the chaoticity
of the Solar system; then the computational complexity
of the (theoretical) three-body problem is considered; then
for a special artificial case of this problem (Sitnikov problem)
we show that no analytical theory or any other kind of $O(\P(\log t))$
computation for for that particular problem can exist.

\section*{Chaos in the Solar system}

Analyzing the presence of chaos in the planetary motion has
always been a hard piece of work. At the end of 18-th
century, Laplace and Lagrange proved that (their mathematical
model of) the Solar system is stable. Later,  Le Verrier showed
that high-order terms of the expansion of Newtonian laws,
discarded in their proof, \textit{can} completely change
the picture on a large interval of time. However, the question
of whether the Solar system is \textit{actually} stable or chaotic
remains unsolved from the theoretical point of view.

In the computer era, numerical experiments have been performed by
various researches to estimate the Lyapunov exponent of
planetary orbits, and hence their chaoticity. Laskar (1994)
found that the inner planets of the Solar system have
chaotic behavior and the outer ones do not. 
Sussman and Wisdom (1992) ran their own simulations
that showed the chaotic behavior of the outer planets.
Hayes (2007) has demonstrated that the outer planets may show
chaotic or stable behavior depending on initial conditions,
in both cases within the observational accuracy.

From now on, we shift from the Solar system with its
uncertain parameters, incomplete models and often
error-prone numerical integration methods, to the three
body problem in its pure mathematical formulation.

\section*{The computational complexity of a dynamical system}

We study the computational complexity of the initial value problem (IVP) for
the dynamical system

\begin{equation}\label{ode}
\begin{array}{lcl}
  \dot \xvec & = & \fvec(\xvec) \\
  \xvec(0) & = & \xvec_0
\end{array}
\end{equation}

\noindent where $\xvec \in D$, $\xvec_0 \in D$ is a real vector, and
$\fvec: D \to \R^n$ is a computable real vector-valued function
(open set $D\subseteq \R^n$ is the phase space of the system).
We deal with the case when the solution $\xvec^*(t) : \R \rightarrow D$:
(i) exists on the whole $\R$; (ii) is unique; (iii) is a computable real
vector-valued function.

The input data for a problem will be:
the initial conditions $\xvec_0$, the point $t$ in time,
and the precision $\varepsilon$.
A Turing machine TM implementing a (numerically integrated or other)
solution is supposed to consume the input and produce
the output---an approximate state of the system at time $t$---that
matches the actual state of the system up to $\varepsilon$.
$t$ and $\varepsilon$ a treated as rationals, while the
initial conditions are provided
by an oracle  $\varphi$
that gives increasingly precise numbers on demand (Kawamura, 2014).

\begin{Def}\label{ivpfunc}
The solution function of an initial value problem (\ref{ode}) is
the function $S(\xvec_0, t): D \times \Q \to D$, where $S|_{\xvec=\xvec_0}: \Q \to D$ is a computable real vector-valued function, whose closure on the real
axis is the solution of (\ref{ode}).
\end{Def}

\begin{Def}\label{turing}
  Turing machine that computes the solution function of an IVP is
  a Turing machine that accepts rational $t$ and $\varepsilon$ as input;
  has an oracle $\varphi$ that instruments $\xvec_0$ as a computable
  real vector; and produces the value of the solution $\xvec(t)$ corresponding
  to given $\xvec_0$ and $t$, with the precision $\varepsilon$.
\end{Def}

\begin{center}
\includegraphics[width=0.9\linewidth]{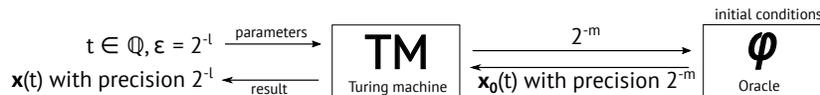}
\captionof{figure}{ Turing machine computing the state of a dynamical system}
\end{center}

We say that an IVP has \textsl{polynomial complexity} if there exists a Turing
machine that computes its solution function in time bounded by $\P(\LENGTH(t),
\LENGTH(\varepsilon))$.

\section*{Computational complexity of the IVP for the $N$-body problem}

Gravitational $N$-body problem is concerned with the Newtonian motion of $N$
point-masses in three dimensions. The system of ODEs for this problem is
the following:
\begin{equation}\label{nbody}
\left. \begin{array}{rcl}
 \dot \pvec_i & = & \vvec_i, \quad i = 1..N\\
 \dot \vvec_i & = & \sum\limits_{\substack{j=1\\j\neq i}}^N\mu_j \frac{\pvec_j-\pvec_i}{|\pvec_j-\pvec_i|^3}, \quad i = 1..N
\end{array} \quad \right\} 
\end{equation}
where $\mu_i\in\R$, $\mu_i \geq 0$, $\pvec_i \in \R^3$, $\vvec_i \in \R^3$.

The initial state of the system is given by a $(7N)$-vector
$$\xvec_0=(\mu_1,\mu_2,\mu_3,p_{1,1},\ldots,p_{N,3},v_{1,1},\ldots,v_{N,3}),$$
while the system (\ref{nbody}) defines a computable real vector-valued
function $\dot\xvec = {\mathbf f}(\xvec)$.

With $N=2$, there is an algebraic solution involving $t$ as a parameter
to a periodic function, and it is not hard to show that the IVP for the
two body problem has polynomial complexity (if the solution exists).

With $N=3$, as shown by Poincar\'e, no algebraic solution exists.
Sundman (1912), however, derived a solution in the form of
converging series. Unfortunately, the estimate of the number of terms
required to calculate the series at point $t$ with a sensible precision
is exponential in $t$ (Belorizky, 1930). 

There has been developed no series (or, generally speaking, no algorithm) able
to calculate the state of a three-body system in a time polynomial in
$\log t$. Intuitively, that is expectable: in a system with a high sensitivity to
initial conditions (i.e. nonzero Lyapunov exponent), any algorithm must consume
at least $O(t)$ input data, otherwise it does not know the initial state with
enough precision to calculate the answer. Since $O(t)$ dominates any
$\P(\log t)$, a polynomial algorithm must not be possible.

What do we really know about the sensitivity to initial conditions in
three body systems? Poincar\'e's discovery was qualitative and did
not include a measure of sensitivity. A number of papers exist with
numerical estimation of Lyapunov exponents (Froeschl\'e 1970,
Quarles et al. 2011), but no theoretical lower bound of the Lyapunov
exponent value is known (however, there is a work by Shevchenko (2004)
with analytical upper bounds).

\medskip

\begin{center}
\def\arraystretch{1.2}
\setlength\tabcolsep{0.5cm}
\begin{tabular}{ p{5cm}  p{5cm} }
\toprule
\textbf{Two body problem} & \textbf{General Case of Three body problem}\\
\midrule
Two-dimensional phase space & Phase space of three or more dimensions \\
\midrule
Algebraic solution known & Algebraic solution proven not to exist \\
\midrule
Regular orbits & Chaotic orbits \\
\midrule
Zero Lyapunov exponent & Supposedly nonzero Lyapunov exponent \\
\midrule
Orbits in practice are calculated in $O(\LENGTH(t)^2)$ time using trivial arithmetical operations, trigonometric functions and Kepler equation solving &
Orbits in practice are calculated in $O(t)$ time using numerical integrators \\
\bottomrule
\end{tabular}
\captionof{table}{ Fundamental differences between two body and three body problems---all related to each other}
\label{table23}\end{center}

The objectives of this work is to strengthen the link between
the chaoticity of dynamical systems and computational complexity,
and strictly prove the computational complexity of a particular
case of the three body problem.

\section*{Sitnikov problem}
From now on, we will focus on a special case of the three-body problem
probably first proposed by Andrei Kolmogorov to Kirill Sitnikov, who was
his student at the time.

{
  \centering
\includegraphics[width=10cm]{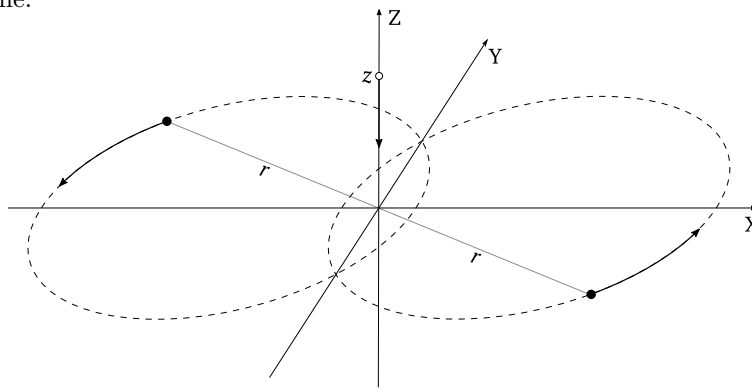}
\captionof{figure}{ Sitnikov problem}
\label{fig-sitnikov}
}

In this problem, two of the bodies are of equal positive mass, while the third body
is massless and lies on a line perpendicular to the plane of the motion
of the first two bodies and passing through their center of mass 
(Fig.~\ref{fig-sitnikov}). Hence, the two bodies follow the unperturbed
(Keplerian) orbit; in this problem, the elliptic orbit is the case.

Let us place the center of mass at the origin, and the $Z$ axis along
the line where the third body is. Let us denote $r(t)$ the distance
from the first body (and the second, as their trajectories are symmetric)
to the origin.

Following Newtonian laws (\ref{nbody}), the coordinate of the third body,
denoted as $z$, obeys the following differential equation:
\begin{equation}\label{sitnikov}
\ddot z = -\frac{2 \mu z}{\sqrt{z^2 + r(t)^2}^3},
\end{equation}

\noindent where $\mu$ is the gravitational constant of the first and second bodies.
Periodic function $r(t)$ comes from the elliptical solution of the two-body problem:

\begin{equation}\label{twobody}
 \begin{array}{rcl}
   r(t) &=& a(1-e \cos E(t)) \\ 
   E(t) - e \sin E(t) &=& \sqrt\frac{2\mu}{a^3}(t - t_0)
   \end{array}
\end{equation}

Semimajor axis $a$, eccentricity $e$, and epoch $t_0$ are constants that
can be calculated from the initial state of the two bodies. $E(t)$ is the
eccentric anomaly angle. The period of $r(t)$ is $P = 2\pi \sqrt\frac{a^3}{2\mu}$.

The initial values in the Sitnikov problem are:

\begin{enumerate}[$\quad\triangleright\quad$]
\item $a > 0$, $e \in (0..1)$, $\mu > 0$ --- parameters of the orbit of the two bodies;
\item $z_0 = z(0)$ --- initial position of the third body in the $Z$ axis.
\item $v_0 = \dot z(0)$ --- initial velocity of the third body in the $Z$ axis.
\item $\phi = E(0)$, $0 \leq \phi < 2 \pi$  --- initial value of the eccentric anomaly of
  the orbit of the two bodies.
  \end{enumerate}
 
The state vector of the system is accordingly
$\xvec = (a,e,\mu,z,v,E)$. $a$, $e$ and $\mu$ do not depend on time;
$\dot z = v$; $\dot v = \ddot z$ from (\ref{sitnikov});
$\dot E$ follows from (\ref{twobody}):
\begin{equation}\label{sitnikov-xdot}
\begin{array}{rcl}
\dot \xvec & = & \fvec(\xvec) = (0, 0, 0, v, \ddot z, \dot E) \\
\ddot z & = & -\frac{2\mu z}{\sqrt{z^2+a^2(1 - e\cos E)^2}^3} \\
\dot E & = & \frac{\sqrt{2\mu}}{\sqrt{a}^3(1 - e \cos E)}
\end{array}
\end{equation}

This very simple system became a rare example of
mathematically proven chaos in gravitational dynamics.
Sitnikov (1961) showed the existence of oscillatory trajectories.
Alexeyev (1968--1969) significantly extended his result, not only
discovering the existence of \textit{all} the classes of final
motions in this problem, but developing the whole theory
of symbolic dynamics for it, and finally proving the following:

\begin{Thm}
  For any sufficiently small eccentricity $e > 0$ there exists an $m(e)$ such that
  for any double-infinite sequence $\{s_n\}_{n\in\Z}, s_n \geq m$
  there exists a solution $z(t)$ of the equation (\ref{sitnikov})
  whose roots satisfy the equation
\begin{equation}\label{sktau}
\left\lfloor \frac{\tau_{k+1}-\tau_k}{P}\right\rfloor = s_k,\ \forall k \in \Z.
\end{equation}
\end{Thm}

\begin{center}
  \includegraphics[width=0.9\linewidth]{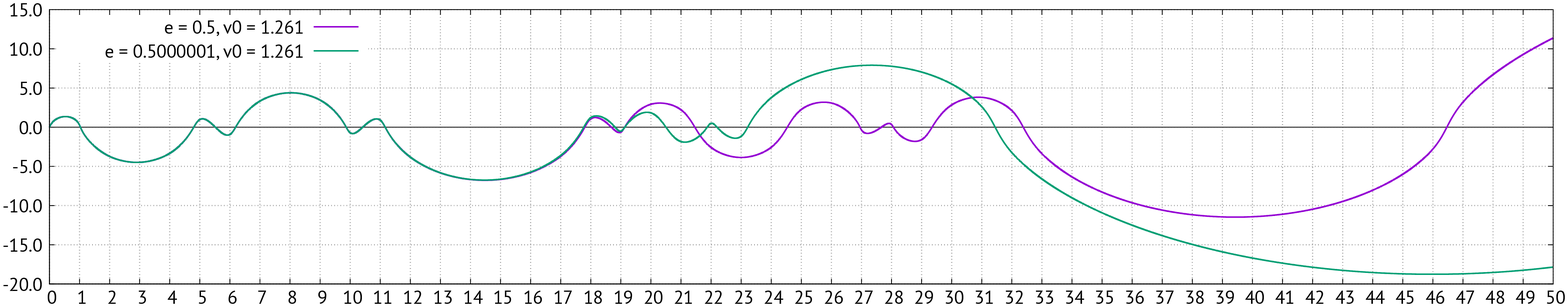}
  \includegraphics[width=0.9\linewidth]{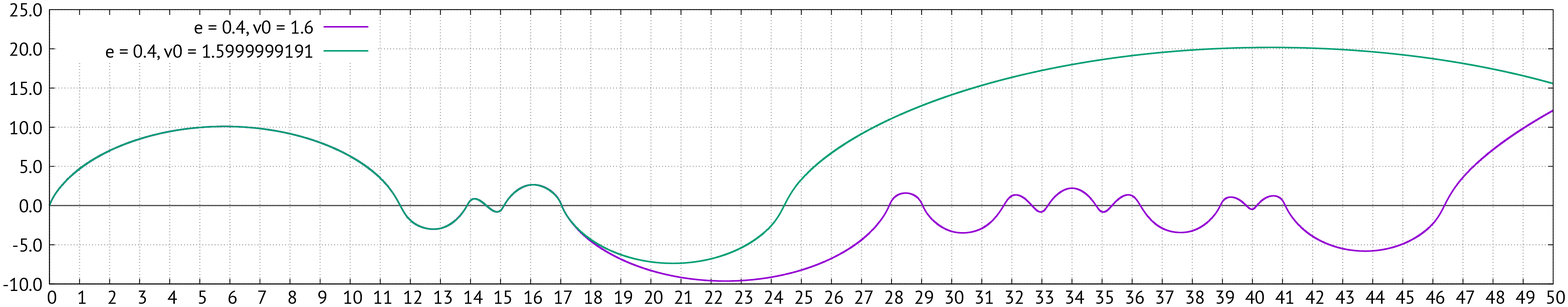}
\captionof{figure}{Different orbits obtained in the Sitnikov problem by small variations of initial parameters. Plots of $z(\frac{t}{2\pi})$ are drawn, showing high sensitivity to parameters $e$ and $v_0$. $z(0) = 0, a = 1, \mu = 0.5$.}
\end{center}

A shortened version of the original theorem is given, excluding the
finite and semi-infinite sequences. Alexeyev also proved a generalization
of his theorem to the case when the third body has a nonzero mass.
A simpler proof was later obtained by Moser (1973).

\section*{Computational complexity of the IVP for the Sitnikov problem}
In (Vasiliev and Pavlov, 2016), it is shown that the solution
of the IVP for the Sitnikov problem always exist (i.e. does not have
singularities), is unique and computable. Then, the following theorem
is proven:
\begin{Thm}\label{main}
  The time complexity of an initial value problem for the Sitnikov problem
  with any fixed value of eccentricity does not have a polynomial upper bound.
\end{Thm}
In other words, no Turing machine can
compute the solution function in time polynomial in $\LENGTH(t)$.
The proof is based on the fact such a Turing machine
also could \textit{restore} a sequence (\ref{sktau})
from Alexeyev's theorem---any sequence from an exponential (in $t$)
number of sequences---reading no more than $\P(\LENGTH(t))$ input
data, which is impossible. Lyapunov exponent is not derived nor used in the proof.

Since the Sitnikov problem can be (in polynomial time) reduced to
the general three-body problem, the above result means that there
can not be a polynomial-time Turing machine for the latter, too.

\section*{Upper bound of computational complexity}
After proving the lower ($O(t)$) bound for the computational complexity,
it is natural to wonder whether the IVP for three body problem
is actually solvable in $O(t)$ (or, even, solvable at all).
Numerical integrators widely used in practice, though run in
$O(t)$ time, rarely \textit{compute} the solution
function with arbitrary precision as it is required by
the definition. The vast majority of these integrators suffer
from \textsl{saturation}:
the step size being small enough, the error grows upon further
decrease of the step size. Therefore, these
integrators can not in principle obtain a solution up to an arbitrary
precision.

One exception is a a modification of Picard--Lindel\"of method
(Matculevich et al. 2013) that computes the solution function
(regardless of how much time it needs) if $\fvec$
is Lipschitz-continuous on $D$.

More recently (Pouly and Gra\c{c}a, 2016), an algorithm
was given for computing the solution function of a polynomial
ODE in a time polynomial in $\LENGTH(\varepsilon)$ and the
length of the solution curve from $0$ to $t$---which, in
many cases of the problem, is linearly bounded by $t$.

\section*{Implications from computational complexity bounds}
The proven non-polynomiality is a theoretical asymptotic statement.
Even for the Sitnikov problem, there are not yet known obstacles to develop an
analytical theory with a fixed but practically sufficient precision available
for a sufficiently large period of time. For other cases of the three body
problem, or for the Solar system, there may exist yet another special
techniques.

\bigskip

{
Numerical theories are irreplaceable--—for a specially constructed dynamical
system and for unlimited requirements for the precision and the time span. There
can not exist quickly converging series for that case, and hence for the general
case. For other particular cases, including the Solar system, we do not know
yet.
}

\section*{Connection between computational complexity and dynamical chaos}

Looking at Table~\ref{table23}, and adding into it a new
partition ``polynomial -- not polynomial'' one could suggest
a link between integrability, computational complexity,
Lyapunov exponents, and chaotic orbits. 
The choice of the three-body problem and oscillatory trajectories is not
principal. We believe that similar results can be obtained in other systems,
where, with the help of methods of symbolic dynamics, complex dynamical behavior
can be shown and analyzed.

For the integrable dynamical systems---those who
have computable integrals of motion with good complexity bounds in $t$ and
$\varepsilon$---it is possible to derive complexity bounds for the initial value problem
in our formal statement. Those bounds will be polynomial by
$\log(t)$ and $\log(1/\varepsilon)$. That can point to a link between
computational complexity of the IVP and integrability.

A strong link between computational complexity and dynamical systems is
known: Ercsey-Ravasz and Toroczkai (2011) showed that an NP-complete
problem (SAT) can be reduced to a problem of finding an attractor
of a dynamical system. A number or other NP-complete problems has
been since then reduced to finding various properties of dynamical systems
(Ahmadi et al. 2013). However, those strong links are one-way: to our
knowledge, there is no bridge \textit{from} continuous dynamical systems
\textit{to} computational complexity.

\end{document}